\documentstyle[12pt]{article}
\begin{document}

\def\ov{\overline}
\def\var{\varepsilon}
\def\om{\omega}
\def\pa{\partial}
\def\la{\langle}
\def\ra{\rangle}
\def\om{\omega}

\title{
Quantum decoherence and the Glauber dynamics\\
from the Stochastic limit
}
\author{L. Accardi, S.V. Kozyrev}
\maketitle

\centerline{\it Centro Vito Volterra Universita di Roma Tor Vergata}

\begin{abstract}
The effects of decoherence for quantum system coupled with a bosonic
field are investigated. An application of the stochastic golden rule
shows that in the stochastic limit the dynamics of such a system
is described by a quantum stochastic differential equation.

The corresponding master equation describes convergence
of a system to equilibrium.
In particular it predicts exponential damping for off--diagonal
matrix elements of the system density matrix, moreover these elements
for a generic system will decay at least as $\exp(-tN{kT\over\hbar})$,
where $N$ is a number of particles in the system.

As an application of the described technique a derivation from first
principles (i.e. starting from a Hamiltonian description) of a quantum
extension of the Glauber dynamics for systems of spins is given.
\end{abstract}

\section{Introduction}

In the present paper we investigate a general model of quantum
system interacting with a bosonic reservoir via an Hamiltonian of the form
$$
H=H_0+\lambda H_I
$$
where $H_0$ is called the free Hamiltonian and $H_I$ the interaction
Hamiltonian.

The stochastic golden rules, which arise in the stochastic limit of
quantum theory as natural generalizations of Fermi golden rule
\cite{[2]}, \cite{[1]},
provide a natural tool to associate a stochastic flow,  driven by a
white noise equation (stochastic Schr\"odinger) equation, to any discrete
system interacting with a quantum field.
This white noise Hamiltonian equation which, when put in normal order
becomes equivalent to a quantum stochastic differential equation.
The Langevin (stochastic Heisenberg) and master equations
are deduced from this white noise equation by means of standard procedures
which are described in \cite{[2]}.

We use these equations to investigate the decoherence in quantum systems.

In the work \cite{[4]}, extending previous results obtained with
perturbative techniques by \cite{CaLeg}, it was shown on the example of the
spin--boson Hamiltonian that
the decoherence in quantum systems can be controlled by the following
constants
(cf. section 2 for the definition of the quantities involved)
$$
\hbox{Re}\, (g|g)^+_{\omega}=\int dk\, |g(k)|^2 2\pi
\delta(\omega(k)-\omega) N(k),
$$
For the simplest case of the equalibrium state of the reservoir
with the temperature $\beta^{-1}=kT$ this constant will be equal to
${kT\over\hbar}$ (actually this is true for large temperatures
and for the dispersion function $\omega(k)=|k|$).

In this paper we extend the approach of \cite{[4]} from $2$--level
systems to arbitrary quantum systems with discrete spectrum. Our resuts show
that the stocastic limit technique gives us an effective method to control
quantum decoherence.

We find that, under the above mentioned interaction, all the off--diagonal
matrix elements, of the density matrix of a generic discrete quantum system,
will decay exponentially if $\hbox{Re}\, (g|g)$ are nonzero.
In other words we obtain the asymptotic diagonalization of the density
matrix.

Moreover, we show that for generic quantum system
the off--diagonal elements of the density matrix decay exponentially
as $\exp(-N\hbox{Re}\, (g|g)t)$,
with the exponent proportional to the number $N$ of particles in the system.
Therefore for generic macroscopic (large $N$) systems the quantum
state will collapse into the classical state very quickly. This effect
was built in {\it by hands} in several phenomenological models of the
quantum measurement process. In the stochastic limit approach it is
deduced from the Hamiltonian model.

This observation contributes to the clarification of one of the old problems
of quantum theory: Why macroscopic systems usually behave classically?
i.e. why do we observe classical states although the evolution of the system
is a unitary operator described by the Shr\"odinger equation?

Moreover, this result allows to distinguish between macroscopic systems
(that behave classically) and microscopic systems (where quantum effects
are important).
Quantum effects (or effects of quantum interference) are connected with
the off--diagonal elements of the density matrix.
Therefore the following notion is natural:
the macroscopic system is a system where off--diagonal elements
of the density matrix decay quickly
(faster than the minimal time of observation).
Using that off--diagonal elements decay as $\exp(-N\hbox{Re}\, (g|g)t)$,
we get the following definition of the macroscopic system:
$N\hbox{Re}\, (g|g)>>1$.

The quantum Markov semigroup we obtain lives invariant the algebra of
the spectral projections of the system Hamiltonian and the associated
master equation, when restricted to the diagonal part of the density
matrix, takes the form of a standard classical kinetic equation,
describing the convergence to equilibrium (Gibbs state) of the system,
coupled with the given reservoir (quantum field).

Summing up: the convergence to equilibrium is a result of quantum
decoherence.

If we can control the interaction so that some of the constants
$\hbox{Re}\, (g|g)$ are zero, then the corresponding matrix elements will
not decay in the stochastic approximation, i.e. in a time scale which is
extremely long with respect to the {\it slow clock} of the discrete system.
In this sense the stochastic limit approach provides a method
for controlling quantum coherence.

The general idea of the stochastic limit (see \cite{[2]}) is to
make the time rescaling $t\to t/\lambda^2$ in the solution of the
Schr\"odinger (or Heisenberg) equation in interaction picture
$U^{(\lambda)}_t=e^{itH_0}e^{-itH}$, associated to the Hamiltonian $H$,
i.e.
$$
{\partial\over\partial t}\,U^{(\lambda)}_t=-i\lambda H_I(t)
U^{(\lambda)}_t
$$
with $H_I(t)=e^{itH_0}H_Ie^{-itH_0}$. This gives the rescaled equation
\begin{equation}\label{1}
{\partial\over\partial t}\,U^{(\lambda)}_{t/\lambda^2}=-{i\over
\lambda}\,H_I(t/\lambda^2)U^{(\lambda)}_{t/\lambda^2}
\end{equation}
and one wants to study the limits, in a topology to be specified,
\begin{equation}\label{2}
\lim_{\lambda\to0}U^{(\lambda)}_{t/\lambda^2}=U_t\ ;\quad
\lim_{\lambda\to0}{1\over\lambda}\,H_I\left({t\over\lambda^2}
\right)=H_t
\end{equation}
The limit $\lambda\to0$
after the rescaling $t\to t/\lambda^2$ is equivalent to the simultaneous
limit $\lambda\to0$, $t\to\infty$ under the condition that $\lambda^2t$
tends to a constant (interpreted as a new slow time scale). This limit
captures the dominating contributions to the dynamics, in a regime of
long times and small coupling, arising from the cumulative effects, on a
large time scale, of small interactions $(\lambda\to0)$. The physical
idea is that, looked from the slow time scale of the atom, the field
looks like a very chaotic object: a {\it quantum white noise\/}, i.e. a
$\delta$--correlated (in time) quantum field $b^*(t,k)$, $b(t,k)$ also
called a {\it master field\/}.

The structure of the present paper is as follows.

In section 2 we introduce the model and consider its stochastic limit.

In section 3 we derive the Langevin equation.

In section 4 we derive the master equation for the density matrix
and show that for non--zero decoherence the master equation describes
the collapse of the density matrix to the classical Gibbs distribution
and discuss the connection of this fact with the procedure of quantum
measurement.

In section 5, using the characterization of quantum decoherence obtained in
section 4 and generalizing arguments of \cite{[4]}, we find that our general
model exhibits macroscopic quantum effects (in particular, conservation
of quantum coherence). These effects are controllable by the state of
the reservoir (that can be controlled by filtering).

In section 6 we apply our general scheme to the
model of a quantum system of spins interacting with bosonic field
and derive a quantum extension of the Glauber dynamics.

Thus our stochastic limit approach provides a microscopic interpretation,
in terms of fundamental Hamiltonian models, to the dynamics of quantum spin
systems. Moreover we deduce the full stochastic equation and not only
the master equation. This is new even in the case of classical spin systems.

\section{The model and it's stochastic limit}

In the present paper we consider a general model, describing the
interaction of a system $S$ with a reservoir, represented by a bosonic
quantum field. Particular cases of this general model were
investigated in \cite{[4]}, \cite{[5]}, \cite{[6]}.
The total Hamiltonian is
$$
H=H_0+\lambda H_I=H_S+H_R+\lambda H_I$$
where $H_R$ is the free Hamiltonian of a bosonic reservoir $R$:
$$H_R=\int\omega(k)a^*(k)a(k)dk$$
acting in the representation space ${\cal F}$ corresponding to the
state $\langle\cdot\rangle$ of bosonic reservoir
generated by the density matrix ${\bf N}$ that we take in the algebra
of spectral projections of the reservoir Hamiltonian.
The reference
state $\langle\cdot\rangle$ of the field is a mean zero
gauge invariant Gaussian state, characterized by
the second order correlation function equal to
$$
\langle a(k)a^*(k')\rangle=(N(k)+1)\delta(k-k')
$$
$$
\langle a^*(k)a(k')\rangle=N(k)\delta(k-k')
$$
where the function $N(k)$ describes the density of bosons with frequency
$k$.
One of the examples is the (gaussian)
bosonic equilibrium state at temperature $\beta^{-1}$.

The system Hamiltonian has the following spectral decomposition
$$
H_S=\sum_r \varepsilon_r P_{\varepsilon_r}
$$
where the index $r$ labels the spectral projections of $H_S$.
For example, for a non--degenerate eigenvalue
$\varepsilon_r$ of $H_S$ the corresponding spectral projection is
$$
P_{\varepsilon_r}= |\varepsilon_r\rangle\langle\varepsilon_r|
$$
where $|\varepsilon_r\rangle$ is the corresponding eigenvector.

The interaction Hamiltonian $H_I$ (acting in ${\cal H}_S\otimes
{\cal F}$) has the form
$$
H_I=\sum_j \left(D_j^*\otimes A(g_j)+D_j\otimes A^*(g_j)\right),
\quad A(g)=\int dk\overline g(k)a(k)\ ,
$$
where $A(g)$ is a smeared quantum field with cutoff function (form
factor) $g(k)$. To perform the construction of the stochastic limit one
needs to calculate the free evolution of the interaction Hamiltonian:
$H_I(t)=e^{itH_0}H_Ie^{-itH_0}$.

Using the identity
$$
1=\sum_r P_{\varepsilon_r}
$$
we write the interaction Hamiltonian in the form
\begin{equation}\label{3}
H_I=\sum_j \sum_{rr'}
P_{\varepsilon_r}D_j^*P_{\varepsilon_r'}\int dk\overline g_j(k)a(k)+h.c.
\end{equation}
Let us introduce the set of energy differences (Bohr frequencies)
$$
F = \{\omega=\varepsilon_r-\varepsilon_{r'}:
\varepsilon_{r}, \varepsilon_{r'}\in \hbox{Spec}\,H_S\}
$$
and the set of all energies of the form
$$
F_{\omega}=\{\varepsilon_{r}:
\exists\varepsilon_{r'}
~(\varepsilon_{r},\varepsilon_{r'}\in \hbox{Spec}\,H_S) \hbox{ such that }
\varepsilon_{r}-\varepsilon_{r'}=\omega  \}
$$
With these notations we rewrite the interaction Hamiltonian (\ref{3})
in the form
$$
H_I=
\sum_j \sum_{\omega\in F} \sum_{\varepsilon_{r}\in F_{\omega}}
P_{\varepsilon_r}D_j^*P_{\varepsilon_{r}-\omega}\int dk\overline g_j(k)
a(k)+h.c.=
$$
\begin{equation}\label{HI}
=\sum_j \sum_{\omega\in F}
E_{\omega}^*\left(D_j\right)
\int dk\overline g_j(k) a(k)+h.c.
\end{equation}
where
\begin{equation}\label{E}
E_{\omega}(X):=
\sum_{\varepsilon_{r}\in F_{\omega}}
P_{\varepsilon_r-\omega} X P_{\varepsilon_{r}}
\end{equation}
It is easy to see that the free volution of $E_{\omega}(X)$ is
$$
e^{itH_S}E_{\omega}(X)e^{-itH_S}=e^{-it\omega}E_{\omega}(X)
$$
Using the formula for the free evolution of bosonic fields
$$
e^{itH_R}a(k)e^{-itH_R}=e^{-it\omega(k)}a(k)
$$
we get for the free evolution of the interaction Hamiltonian:
\begin{equation}\label{HIt}
H_I(t)=\sum_j \sum_{\omega\in F}
E_{\omega}^*\left(D_j\right)
\int dk\overline g_j(k) e^{-it(\omega(k)-\omega)}a(k)+h.c.
\end{equation}
In the stochastic limit the field $H_I(t)$ gives rise to a family of
quantum white noises, or master fields. To investigate these noises, let
us suppose the following:

1) $\omega(k)\geq0$, $\forall\,k$;

2) The $d-1$--dimensional Lebesgue measure of the surface $\{k:\omega
(k)=0\}$ is equal to zero (so that $\delta(\omega(k))=0$) (for example
$\omega(k)=k^2+m$ with $m\geq0$).

Now let us investigate the limit of $H_I(t/\lambda^2)$ using one of the
basic formulae of the stochastic limit:
\begin{equation}\label{4}
\lim_{\lambda\to0}{1\over\lambda^2}\,\exp\left({it\over\lambda^2}\,
f(k)\right)=2\pi\delta(t)\delta(f(k))
\end{equation}
which shows that the term $\delta(f(k))$ in (\ref{4}) is not identically
equal to zero only if $f(k)=0$ for some $k$ in a set of nonzero
$d-1$--dimensional Lebesgue measure. This explains condition (2) above.

The rescaled interaction Hamiltonian is expressed in terms of the rescaled
creation and annihilation operators
$$
a_{\lambda,\omega}(t,k)=
{1\over\lambda}\,e^{-i{t\over\lambda^2}\,(\omega(k)-\omega)}a(k),\quad
\omega \in F
$$
After the stochastic limit every rescaled annihilation
operator corresponding to any transition from
$\varepsilon_{r'}$ to $\varepsilon_{r}$ with the frequency
$\omega=\varepsilon_{r}-\varepsilon_{r'}$
generates one non--trivial quantum white noise
$$
b_{\omega}(t,k)=\lim_{\lambda\to0}a_{\lambda,\omega}
(t,k)=\lim_{\lambda\to0}{1\over\lambda}\,e^{-i{t\over\lambda^2}\,
(\omega(k)-\omega)}a(k)
$$
with the relations
$$[b_{\omega}(t,k),b^*_{\omega}(t',k')]=
\lim_{\lambda\to0}[a_{\lambda,\omega}(t,k),a^*_{\lambda,
\omega}(t',k')]=
$$
\begin{equation}\label{5}
=\lim_{\lambda\to0}{1\over\lambda^2}\,e^{-i{t-t'\over\lambda^2}\,
(\omega(k)-\omega)}\delta(k-k')=
2\pi\delta(t-t')
\delta(\omega(k)-\omega)\delta(k-k')
\end{equation}
$$
[b_{\omega}(t,k),b^*_{\omega'}(t',k')]=0
$$
(cf. \ref{4}). This shows, in particular that quantum white noises,
corresponding to different Bohr frequencies, are mutually independent.

The stochastic limit of the interaction Hamiltonian is therefore equal to
\begin{equation}\label{6}
h(t)
=\sum_j \sum_{\omega\in F} E_{\omega}^*\left(D_j\right)
\int dk\overline g_j(k) b_{\omega}(t,k)+h.c.
\end{equation}

The state of the master field (white noise) $b_{\omega}(t,k)$,
corresponding to our choice of the initial state of the field, is the mean
zero gauge invariant Gaussian state with correlations:
$$
\langle b^*_{\omega}(t,k)b_{\omega}(t',k')\rangle=
2\pi\delta(t-t')\delta(\omega(k)-\omega)\delta(k-k')N(k)
$$
$$
\langle b_{\omega}(t,k)b^*_{\omega}(t',k')\rangle=
2\pi\delta(t-t')\delta(\omega(k)-\omega)\delta(k-k')(N(k)+1)
$$
and vanishes for noises corresponding to different Bohr frequences.

Now let us investigate the evolution equation in interaction picture for
our model. According to the general scheme of the stochastic limit, we
get the (singular) white noise equation
\begin{equation}\label{7}
{d\over dt}U_t=-ih(t)U_t
\end{equation}
whose normally ordered form is
the quantum stochastic differential equation \cite{[3]}
\begin{equation}\label{8}
dU_t=(-idH(t)-Gdt)U_t
\end{equation}
where $h(t)$ is the white noise Hamiltonian
(\ref{6}) given by the stochastic limit
of the interaction Hamiltonian and
\begin{equation}\label{9}
dH(t)=
\sum_j \sum_{\omega\in F}
\left(
E_{\omega}^*\left(D_j\right)dB_{j\omega}(t)+
E_{\omega}\left(D_j\right)dB^*_{j\omega}(t)
\right)
\end{equation}
\begin{equation}\label{10}
dB_{j\omega}(t)=\int dk\overline g_j(k)\int^{t+dt}_tb_{\omega}(\tau,k)d\tau
\end{equation}

According to the stochastic golden rule (\ref{8}) the limit dynamical
equation is obtained as follows:
the first term in (\ref{8}) is just the limit of the iterated series
solution for (\ref{1})
$$
\lim_{\lambda\to0}{1\over\lambda}\,\int^{t+dt}_t
H_I\left({\tau\over \lambda^2}\right)d\tau
$$
The second term $Gdt$, called the drift, is equal to the limit of the
expectation value in the reservoir state of the second term in the
iterated series solution for (\ref{1})
$$
\lim_{\lambda\to0}{1\over\lambda^2}\,
\int^{t+dt}_t dt_1 \int^{t_1}_t dt_2
\langle H_I\left({t_1\over \lambda^2}\right)
H_I\left({t_2\over\lambda^2}\right)\rangle
$$
Making in this formula the change of variables $\tau=t_2-t_1$ we get
\begin{equation}\label{11}
\lim_{\lambda\to0}{1\over\lambda^2}\,\int^{t+dt}_tdt_1\int^0_{t-
t_1}d\tau\langle H_I\left({t_1\over\lambda^2}\right)H_I
\left({t_1\over\lambda^2}\,+{\tau\over\lambda^2}\right)\rangle
\end{equation}
Computing the expectation value and using
the fact that the limits of oscillating factors of the form
$\lim\limits_{\lambda\to0}e^{ict_1\over\lambda^2}$ vanish unless the
constant $c$ is equal to zero, we see that we can have non--zero limit
only when all oscillating factors of a kind $e^{ict_1\over\lambda^2}$
(with $t_1$) in (\ref{11}) cancel. In conclusion we get
$$
G=\sum_{ij} \sum_{\omega\in F}
\int^0_{-\infty}d\tau
\biggl(
\int dk\, \overline{g_i(k)}g_j(k)
e^{i\tau (\omega(k)-\omega)}(N(k)+1)
E_{\omega}^*\left(D_i\right)E_{\omega}\left(D_j\right)+
$$
$$
+\int dk\, {g_i(k)}\overline{g_j(k)}
e^{-i\tau(\omega(k)-\omega)}N(k)
E_{\omega}\left(D_i\right)E_{\omega}^*\left(D_j\right)
\biggr)
$$
and therefore, from the formula
\begin{equation}\label{12}
\int^0_{-\infty}e^{it\omega}dt={-i\over\omega-i0}\,=\pi\delta(\omega)-
i\,\hbox{P.P.}\,{1\over\omega}
\end{equation}
we get the following expression for the drift $G$:
$$
\sum_{ij} \sum_{\omega\in F}
\biggl(
\int dk\, \overline{g_i(k)}g_j(k)
{-i(N(k)+1)\over\omega(k)-\omega-i0}
E_{\omega}^*\left(D_i\right)E_{\omega}\left(D_j\right)+
$$
$$
+
\int dk\, {g_i(k)}\overline{g_j(k)}
{iN(k)\over\omega(k)-\omega+i0}
E_{\omega}\left(D_i\right)E_{\omega}^*\left(D_j\right)
\biggr)=
$$
\begin{equation}\label{13}
=\sum_{ij} \sum_{\omega\in F}
\left(
(g_i|g_j)^-_{\omega}
E_{\omega}^*\left(D_i\right)E_{\omega}\left(D_j\right)
+\overline{(g_i|g_j)}^+_{\omega}
E_{\omega}\left(D_i\right)E_{\omega}^*\left(D_j\right)
\right)
\end{equation}
Let us note that for (\ref{13}) we have the following Cheshire Cat
effect found in \cite{[4]}: even if the frequency $\omega$
is negative and therefore does not generate a quantum white noise
the corresponding values $(g|g)^\pm_{\omega}$ in (\ref{13}) will be
non--zero. In other terms: negative Bohr frequencies contribute to an
energy shift in the system, but not to its damping.
\medskip

\noindent
{\bf Remark}\qquad
If $F$ is any subset of $\hbox{Spec}\, H_S$ and $X_r$ are arbitrary
bounded operators on ${\cal H}_S$ then for any $t\in R$
$$
e^{it H_S} \sum_{\varepsilon_r\in F} P_{\varepsilon_r} X_r
P_{\varepsilon_r}=
\sum_{\varepsilon_r\in F} e^{it \varepsilon_r}
P_{\varepsilon_r} X_r P_{\varepsilon_r}=
\sum_{\varepsilon_r\in F} P_{\varepsilon_r} X_r P_{\varepsilon_r}e^{it H_S}
$$
In other words:
$\sum_{\varepsilon_r\in F} P_{\varepsilon_r} X_r P_{\varepsilon_r}$
belongs to the commutant $L^{\infty}\left(H_S\right)'$
of the abelian algebra $L^{\infty}\left(H_S\right)$,
generated by the spectral projections of $H_S$.

\medskip

A corollary of this remark is that, for each $\omega\in F$, for any bounded
operator $X\in L^{\infty}\left(H_S\right)'$ and for each pair of indices
$(i,j)$ the operators
\begin{equation}\label{exe}
E_{\omega}\left(D_i\right) X E_{\omega}^*\left(D_j\right),\qquad
E_{\omega}^*\left(D_i\right) X E_{\omega}\left(D_j\right)
\end{equation}
belong to the commutant $L^{\infty}\left(H_S\right)'$
of $L^{\infty}\left(H_S\right)$. In particular, if $H_S$ has
non--degenerate spectrum so that $L^{\infty}\left(H_S\right)$
is a maximal abelian subalgebra of $B\left({\cal H}_S\right)$,
the operators (\ref{exe}) also belong to $L^{\infty}\left(H_S\right)$.

\section{The Langevin equation}

Now we will find the Langevin equation, which is the limit of the
Heisenberg evolution, in interaction representation. Let $X$ be an
observable. The Langevin equation is the equation satisfied by the
stochastic flow $j_t$, defined by:
$$
j_t(X)=U^*_t X U_t
$$
where $U_t$ satisfies equation (\ref{8}) in the previous section, i.e.
\begin{equation}\label{14}
dU_t=(-idH(t)-Gdt)U_t
\end{equation}

To derive the Langevin equation we consider
\begin{equation}\label{15}
dj_t(X)=j_{t+dt}(X)-j_t(X)=dU^*_t X U_t+U^*_t X dU_t+
dU^*_t X dU_t
\end{equation}
The only nonvanishing products in the quantum stochastic differentials are
\begin{equation}\label{16}
dB_{i\omega}(t)dB^*_{j\omega}(t)=2\hbox{Re}\,(g_i|g_j)^-_{\omega}dt,\quad
dB^*_{i\omega}(t)dB_{j\omega}(t)=2\hbox{Re}\,(g_i|g_j)^+_{\omega}dt
\end{equation}
Combining the terms in (\ref{15}) and using (\ref{14}), (\ref{9}),
(\ref{13})
and (\ref{16}) we get the Langevin equation
\begin{equation}\label{17}
dj_t(X)=\sum_\alpha j_t\circ\theta_\alpha(X)dM^\alpha(t)
=\sum_{n=-1,1;j\omega}
j_t\circ\theta_{nj\omega}(X)dM^{nj\omega}(t)+j_t\circ\theta_0(X)dt
\end{equation}
where
\begin{equation}\label{18}
dM^{-1,j\omega}(t)=dB_{j\omega}(t),\quad
\theta_{-1,j\omega}(X)=-i[X,E_{\omega}^*\left(D_j\right)]
\end{equation}
\begin{equation}\label{19}
dM^{1,j\omega}(t)=dB^*_{j\omega}(t),\quad
\theta_{1,j\omega}(X)=-i[X,E_{\omega}\left(D_j\right)]
\end{equation}
and
\begin{equation}\label{20}
\theta_0(X)=\sum_{ij} \sum_{\omega\in F}
\biggl(
-i\hbox{ Im }(g_i|g_j)^-_{\omega}
[X,E_{\omega}^*\left(D_i\right)E_{\omega}\left(D_j\right)]+
i\hbox{ Im }{(g_i|g_j)}^+_{\omega}
[X,E_{\omega}\left(D_i\right)E_{\omega}^*\left(D_j\right)]+
$$
$$+
2\hbox{Re}\,(g_i|g_j)^-_{\omega}
\left(
E_{\omega}^*\left(D_i\right) X E_{\omega}\left(D_j\right)
-{1\over 2}
\{X,E_{\omega}^*\left(D_i\right)E_{\omega}\left(D_j\right)\}\right)+
$$
$$
+2\hbox{Re}\,{(g_i|g_j)}^+_{\omega}
\left(
E_{\omega}\left(D_i\right) X E_{\omega}^*\left(D_j\right)
-{1\over 2}
\{X,E_{\omega}\left(D_i\right)E_{\omega}^*\left(D_j\right)\}\right)\biggr)
\end{equation}
is a quantum Markovian generator.
The structure map $\theta_0(X)$ has the standard form of the generator
of a master equation \cite{Lindblad}
$$
\theta_0(X)=\Psi(X)-{1\over2}\{\Psi(1),X\}+i[H,X]
$$
where $\Psi$ is a completely positive map and $H$ is selfadjoint.
In our case $\Psi(X)$ is a linear combination of terms of the type
$$
E_{\omega}^*\left(D_i\right) X E_{\omega}\left(D_j\right)
$$

\medskip

\noindent
{\bf Remark}\qquad
A corollary of the remark at the end of section 2 is that the Markovian
generator $\theta_0$ maps $L^{\infty}(H_S)'$ into itself.
Moreover, if $X$ in (\ref{20}) belongs to the $L^{\infty}(H_S)$
then the Hamiltonian part of $\theta_0(X)$ vanishes and only
the dissipative part remains. In particular, if $H_S$ has non--degenerate
spectrum then $\theta_0(X)$ maps $L^{\infty}(H_S)$ and has the form
$$
\theta_0(X)=\sum_{ij} \sum_{\omega\in F}
\biggl(
2\hbox{Re}\,(g_i|g_j)^-_{\omega}
\left(
E_{\omega}^*\left(D_i\right) X E_{\omega}\left(D_j\right)
-X E_{\omega}^*\left(D_i\right)E_{\omega}\left(D_j\right)\right)+
$$
$$
+2\hbox{Re}\,{(g_i|g_j)}^+_{\omega}
\left(
E_{\omega}\left(D_i\right) X E_{\omega}^*\left(D_j\right)
-X E_{\omega}\left(D_i\right)E_{\omega}^*\left(D_j\right)\right)\biggr)
$$
for any $X\in L^{\infty}(H_S)$.

The structure maps  $\theta_{\alpha}$ in (\ref{17}) satisfy the following
stochastic Leibnitz rule, see the paper \cite{AcKo99a}.
\medskip

\noindent
{\bf Theorem.}\qquad {\sl
For any pair of operators in the system algebra $X$, $Y$, the structure maps
in the Langevin equation (\ref{17}) satisfy the equation
$$
\theta_{\alpha}(XY)=\theta_{\alpha}(X) Y +X\theta_{\alpha}(Y)
+\sum_{\beta,\gamma}c_{\alpha}^{\beta\gamma}
\theta_{\beta}(X)\theta_{\gamma}(Y)
$$
where the structure constants $c_{\alpha}^{\beta\gamma}$
is given by the Ito table
$$
dM^{\beta}(t)dM^{\gamma}(t)=
\sum_{\alpha}c_{\alpha}^{\beta\gamma}dM^{\alpha}(t)
$$
The conjugation rules of $dM^{\alpha}(t)$ and $\theta_{\alpha}$
are connected in such a way that formula (\ref{17})
defines a $*$--flow ($*\circ j_t=j_t\circ *$).
}

\subsection{Evolution for the density matrix}

Let us now investigate the master equation for the density matrix $\rho$.

We will show that if the reservoir
is in the equilibrium state at temperature $\beta^{-1}$
then for the generic system with decoherence
the solution of the master equation $\rho(t)$ with $t\to\infty$
tends to the classical Gibbs state with the same temperature $\beta^{-1}$.
This phenomenon realizes the quantum measurement procedure --- the
quantum state (density matrix) collapses into the classical state.

To show this we use the control of quantum decoherence
that was found in the stochastic approximation of quantum theory,
see \cite{[4]} and discussion below.

Let us consider the evolution of the state (positive normed linear
functional on system observables) given by the density matrix $\rho$,
$\rho(X)=\hbox{tr}\,\hat{\rho} X$.
The evolution of the state is defined as follows
$$
\rho_t=j_t^*(\rho)=\rho\circ j_t
$$
Therefore from (\ref{17}) we get the evolution equation
$$
d\rho_t(X)=\rho\circ dj_t(X)=
\rho\circ\sum_\alpha j_t\circ\theta_\alpha(X)dM^\alpha(t)=
\sum_\alpha\rho_t \left(\theta_\alpha(X)dM^\alpha(t)\right)
$$
Only the stochastic
differential $dt$ in this formula will survive and we get the master
equation
\begin{equation}\label{mas}
{d\over dt}\rho_t(X)=\rho_t\circ\theta_0(X)\equiv  \theta^*_0(\rho_t)(X)
\end{equation}

Let us consider the density
matrix $\hat\rho=\hat\rho_S\otimes \hat\rho_R$,
$$
\hat\rho_{S,t}=\sum_{\mu,\nu}\rho(\mu,\nu,t)|\mu\rangle\langle\nu|
$$
where $|\mu\rangle$, $|\nu\rangle$  are eigenvectors of the system
Hamiltonian $H_S$.

Using the form (\ref{20}) of $\theta_0$
and the identities
$$
\hbox{tr } Y [X,A] =- \hbox{tr } [Y,A] X
$$
$$
\hbox{tr } Y\left(AXB-{1\over 2}\{X,AB\}\right)=
\hbox{tr } \left(BYA-{1\over 2}\{Y,AB\}\right)X
$$
the master equation (\ref{mas})
will take the form
\begin{equation}\label{masterequ}
\sum_{\mu,\nu} {d\over dt}\rho(\mu,\nu,t)|\mu\rangle\langle\nu|=
\sum_{\mu,\nu}\rho(\mu,\nu,t)
$$
$$
\sum_{ij} \sum_{\omega\in F}
\biggl(
i\hbox{ Im }(g_i|g_j)^-_{\omega}
[|\mu\rangle\langle\nu|,
E_{\omega}^*\left(D_i\right)E_{\omega}\left(D_j\right)]
-i\hbox{ Im }{(g_i|g_j)}^+_{\omega}
[|\mu\rangle\langle\nu|,
E_{\omega}\left(D_i\right)E_{\omega}^*\left(D_j\right)]
+$$
$$+
2\hbox{Re}\,(g_i|g_j)^-_{\omega}
\left(
E_{\omega}\left(D_j\right) |\mu\rangle\langle\nu| E_{\omega}^*\left(D_i\right)
-{1\over 2}
\{|\mu\rangle\langle\nu|,
E_{\omega}^*\left(D_i\right)E_{\omega}\left(D_j\right)\}\right)+
$$
$$
+2\hbox{Re}\,{(g_i|g_j)}^+_{\omega}
\left(
E_{\omega}^*\left(D_j\right) |\mu\rangle\langle\nu| E_{\omega}\left(D_i\right)
-{1\over 2}
\{|\mu\rangle\langle\nu|,
E_{\omega}\left(D_i\right)E_{\omega}^*\left(D_j\right)\}\right)\biggr)
=$$
$$=\sum_{\mu,\nu}\rho(\mu,\nu,t)
\sum_{ij} \sum_{\omega\in F}
\biggl(
i\hbox{ Im }(g_i|g_j)^-_{\omega}
\left(
|\mu\rangle\langle\nu|\chi_{\omega}(\varepsilon_{\nu})
D_i^* P_{\varepsilon_{\nu}-\omega} D_j P_{\varepsilon_{\nu}}-
\chi_{\omega}(\varepsilon_{\mu})
P_{\varepsilon_{\mu}} D_i^* P_{\varepsilon_{\mu}-\omega}D_j
|\mu\rangle\langle\nu|
\right)+
$$
$$+
2\hbox{Re}\,(g_i|g_j)^-_{\omega}
\biggl(
\chi_{\omega}(\varepsilon_{\mu})
\chi_{\omega}(\varepsilon_{\nu})
P_{\varepsilon_{\mu}-\omega}D_j
|\mu\rangle\langle\nu|
D_i^*P_{\varepsilon_{\nu}-\omega}-
$$
$$
-{1\over 2}
\left(
|\mu\rangle\langle\nu|\chi_{\omega}(\varepsilon_{\nu})
D_i^* P_{\varepsilon_{\nu}-\omega} D_j P_{\varepsilon_{\nu}}+
\chi_{\omega}(\varepsilon_{\mu})
P_{\varepsilon_{\mu}} D_i^* P_{\varepsilon_{\mu}-\omega}D_j
|\mu\rangle\langle\nu|
\right)
\biggr)-
$$
$$
-i\hbox{ Im }{(g_i|g_j)}^+_{\omega}
\left(
|\mu\rangle\langle\nu|\chi_{-\omega}(\varepsilon_{\nu}+\omega)
D_i P_{\varepsilon_{\nu}+\omega} D_j^* P_{\varepsilon_{\nu}}-
\chi_{-\omega}(\varepsilon_{\mu}+\omega)P_{\varepsilon_{\mu}}
D_i P_{\varepsilon_{\mu}+\omega} D_j^*
|\mu\rangle\langle\nu|
\right)+
$$
$$
+2\hbox{Re}\,{(g_i|g_j)}^+_{\omega}
\biggl(
\chi_{-\omega}(\varepsilon_{\mu}+\omega)
\chi_{-\omega}(\varepsilon_{\nu}+\omega)
P_{\varepsilon_{\mu}+\omega}D_j^*
|\mu\rangle\langle\nu| D_iP_{\varepsilon_{\nu}+\omega}-
$$
$$
-{1\over 2}
\left(
|\mu\rangle\langle\nu|\chi_{-\omega}(\varepsilon_{\nu}+\omega)
D_i P_{\varepsilon_{\nu}+\omega} D_j^* P_{\varepsilon_{\nu}}+
\chi_{-\omega}(\varepsilon_{\mu}+\omega)P_{\varepsilon_{\mu}}
D_i P_{\varepsilon_{\mu}+\omega} D_j^*
|\mu\rangle\langle\nu|
\right)
\biggr)\biggr)
\end{equation}
where $\chi_{\omega}(\varepsilon_{\mu})=1$
if $\varepsilon_{\mu}\in F_{\omega}$ and
equals to 0 otherwise.

\section{Dynamics for generic systems}

Let us investigate the behavior of a system with dynamics
defined by (\ref{masterequ}).
This dynamics will depend on the Hamiltonian of the system.

We will call the Hamiltonian $H_S$ {\it generic\/}, if:

\noindent
1)\qquad
The spectrum $\hbox{ Spec }H_S$ of the Hamiltonian is non degenerate.

\noindent
2)\qquad
For any Bohr frequency $\omega$ there exists a unique pair of energy
levels $\varepsilon$, $\varepsilon'\in \hbox{ Spec }H_S$ such that:
$$
\omega=\varepsilon-\varepsilon'
$$
We investigate (\ref{masterequ}) for generic Hamiltonian.
We also consider the case of one test function $g_i(k)=g(k)$,
although this is not important.
In this case
$$
E_{\omega}(X)=|\sigma'\rangle\langle\sigma'|X|\sigma\rangle\langle\sigma|=
|\sigma'\rangle\langle\sigma| \langle\sigma'|X|\sigma\rangle
$$
where $\omega=\varepsilon_{\sigma}-\varepsilon_{\sigma'}$.
The Markovian generator $\theta_0^{*}$ in (\ref{masterequ}) takes the form
\begin{equation}\label{generic}
\theta_0^{*}(X)=\sum_{\sigma,\sigma'}
|\langle\sigma'|D|\sigma\rangle|^2
\biggl(
i\hbox{ Im }(g|g)^-_{\sigma\sigma'}
[X,|\sigma\rangle\langle\sigma|]+
$$
$$+
2\hbox{Re}\,(g|g)^-_{\sigma\sigma'}
\biggl(
|\sigma'\rangle\langle\sigma'|
\langle\sigma|X|\sigma\rangle -
{1\over 2}
\{X,|\sigma\rangle\langle\sigma|\}\biggr)-
$$
$$
-i\hbox{ Im }(g|g)^+_{\sigma\sigma'}
[X,|\sigma'\rangle\langle\sigma'|]+
2\hbox{Re}\,(g|g)^+_{\sigma \sigma'}
\biggl(
|\sigma\rangle\langle\sigma|
\langle\sigma'|X|\sigma'\rangle-
{1\over 2}
\{X,|\sigma'\rangle\langle\sigma'|\}\biggr)\biggr)
\end{equation}
We use here the notion
$$
(g|g)_{\mu\sigma}=(g|g)_{\varepsilon_{\mu}-\varepsilon_{\sigma}}
$$
Notice that the factors
$\hbox{Re}\,(g|g)^\pm_{\sigma\sigma'}$ are $>0$ only for
$\varepsilon_{\sigma}>\varepsilon_{\sigma'}$ and vanish for
the opposite case.

It is easy to see that the terms in (\ref{generic}) of the form
$$
|\sigma\rangle\langle\sigma|
\langle\sigma'|X|\sigma'\rangle
$$
for off--diagonal elements of the density matrix
$X=|\mu\rangle\langle\nu|$ are equal to zero.
We will show that in such case
the equation (\ref{masterequ}) will predict fast damping
of the states of the kind $|\mu\rangle\langle\nu|$.

In the non--generic case one can expect the fast damping
of the state $|\mu\rangle\langle\nu|$
with different energies
$\varepsilon_{\mu}$ and $\varepsilon_{\nu}$.

With the given assumptions the action of $\theta_0^{*}$ on the off--diagonal
matrix unit $|\mu\rangle\langle\nu|$,
$\varepsilon_{\mu} \ne \varepsilon_{\nu}$
is equal to  $A_{\mu\nu} |\mu\rangle\langle\nu|$ where
the number $A_{\mu\nu}$ is given by the following
\begin{equation}\label{nondiagonal}
A_{\mu\nu}=\sum_{\sigma}\biggl(
i\hbox{Im}\, (g|g)^-_{\mu\sigma}|\langle\sigma|D|\mu\rangle|^2-
i\hbox{Im}\, (g|g)^-_{\nu\sigma} |\langle\sigma|D|\nu\rangle|^2-
i\hbox{Im}\, (g|g)^+_{\sigma\mu}|\langle\mu|D|\sigma\rangle|^2+
$$
$$
+i\hbox{Im}\, (g|g)^+_{\sigma\nu} |\langle\nu|D|\sigma\rangle|^2
-\hbox{Re}\, (g|g)^-_{\mu\sigma}|\langle\sigma|D|\mu\rangle|^2
$$
$$
-\hbox{Re}\, (g|g)^-_{\nu\sigma} |\langle\sigma|D|\nu\rangle|^2
-\hbox{Re}\, (g|g)^+_{\sigma\mu}|\langle\mu|D|\sigma\rangle|^2
-\hbox{Re}\, (g|g)^+_{\sigma\nu} |\langle\nu|D|\sigma\rangle|^2
\biggr)
\end{equation}

The map $\theta^*_0$ multiplies off--diagonal matrix
elements of the density matrix $\hat\rho_S$ by a number $A_{\mu\nu}$.
Let us note that
$$
\hbox{Re}\, A_{\mu\nu} \le 0
$$

Moreover, for generic Hamiltonian
the map $\theta^*_0$ mixes diagonal elements of the density matrix
but does not mix diagonal and off--diagonal elements
(the action
of $\theta^*_0$ on diagonal element is equal to the linear combination
of diagonal elements).

The equation (\ref{masterequ}) for the generic case takes the form
\begin{equation}\label{generic_dm}
\sum_{\mu,\nu} {d\over dt}\rho(\mu,\nu,t)|\mu\rangle\langle\nu|=
\sum_{\mu\ne\nu} A_{\mu\nu}\rho(\mu,\nu,t)|\mu\rangle\langle\nu| +
$$
$$
+\sum_{\sigma} |\sigma\rangle\langle\sigma|
\sum_{\sigma'}\biggl(
\rho(\sigma',t)\left(
2\hbox{Re }(g|g)^-_{\sigma'\sigma} |\langle\sigma|D|\sigma'\rangle|^2
+2\hbox{Re }(g|g)^+_{\sigma\sigma'}|\langle\sigma'|D|\sigma\rangle|^2\right)
-
$$
$$
-\rho(\sigma,t)\left(
2\hbox{Re }(g|g)^+_{\sigma'\sigma}|\langle\sigma|D|\sigma'\rangle|^2
+ 2\hbox{Re }(g|g)^-_{\sigma\sigma'}|\langle\sigma'|D|\sigma\rangle|^2
\right)\biggr)
\end{equation}
with $A_{\mu\nu}$ given by (\ref{nondiagonal}) and
$\rho(\sigma,t)=\rho(\sigma,\sigma,t)$.

For instance we get
$$
j^*_t(|\mu\rangle\langle\nu|)=\exp(A_{\mu\nu}t) |\mu\rangle\langle\nu|
$$
We see that if any of $\hbox{Re}\,(g|g)
|\langle\beta|D|\alpha\rangle|^2$ in (\ref{nondiagonal}) is non--zero
then the corresponding off--diagonal matrix element of the density matrix
decays. We obtain an effect of the diagonalization of the density matrix.
This gives an effective criterium for quantum decoherence in the stochastic
approximation: the system will exhibit decoherence if the constants
$\hbox{Re}\,(g|g)^\pm$ are non--zero.

Now we estimate the velocity of decay of the density matrix
$|\mu\rangle\langle\nu|$ for a quantum system
with $N$ particles. The eigenstate $|\mu\rangle$ of
the Hamiltonian of such a system can be considered as a tensor product
over degrees of freedom of the system of some substates.
Let us estimate from below the number of degrees of freedom
of the system by the number of particles that belong to the system
(for each particle we have few degrees of freedom).
To get the estimate from below for the velocity of decay we
assume that $|\langle\sigma|D|\mu\rangle|^2$ in (\ref{nondiagonal})
is non--zero only if the state $\sigma$ differs from the state $\mu$
only for one degree of freedom.

Then the summation over $\omega$ (or equivalently over $\sigma$)
in (\ref{nondiagonal}) can be estimated by the summation over
the degrees of freedom, or over particles belonging to the system.
If we have total decoherence, i.e. all $\hbox{Re}\, (g|g)$ are non--zero,
then, taking all corresponding $|\langle\sigma|D|\mu\rangle|^2=1$,
we can estimate (\ref{nondiagonal}) as $-N\hbox{Re}\, (g|g)$, where
$N$ is the number of particles in the system, or
\begin{equation}\label{collapse}
j^*_t(|\mu\rangle\langle\nu|)=\exp(-N\hbox{Re}\, (g|g)t)
|\mu\rangle\langle\nu|
\end{equation}
The off--diagonal element of the density matrix decays exponentially,
with the exponent proportional to the number of particles in the system.
Therefore for macroscopic (large $N$) systems with decoherence the quantum
state will collapse into the classical state very quickly.

This observation clarifies,
why macroscopic quantum systems usually behave classically.
The equation (\ref{collapse}) describes such type of behavior,
predicting that the quantum state damps at least as quickly as
$\exp(-N\hbox{Re}\, (g|g)t)$. Therefore a macroscopic system (large $N$)
will become classical in a time of order $(N\hbox{Re}\, (g|g))^{-1}$.

Let us estimate the constant $\hbox{Re}\, (g|g)$ for the equilibrium
state of the reservoir with the temperature $\beta^{-1}=kT$.
In this case
$$
\hbox{Re}\, (g|g)^{-}_{\omega}={1\over\hbar}{\pi\over 1-e^{-\beta \omega}}
\int dk\, |g(k)|^2 \delta(\omega(k)-\omega)
$$
Taking $g(k)=1$ and using that the dispersion function $\omega(k)$
depends only on $|k|$ we get
$$
\int dk\, |g(k)|^2 \delta(\omega(k)-\omega)=
\int d\Omega \int_{0}^{\infty} d\rho \delta(\omega(\rho)-\omega)=
4\pi \int_{0}^{\infty} d\rho \delta(\omega(\rho)-\omega)
$$
where $\int d\Omega$ is the integration over angles. If we take the
dispersion function $\omega(k)=|k|$, then for this integral we get
$4\pi\omega$.

Therefore for this choice of dispersion function we get
$$
\hbox{Re}\, (g|g)^{-}_{\omega} =
{1\over\hbar}{4\pi^2\omega\over 1-e^{-\beta \omega}}
$$
and analogously
$$
\hbox{Re}\, (g|g)^{+}_{\omega} =
{1\over\hbar}{4\pi^2\omega\over e^{\beta \omega}-1}
$$
In the limit of small $\beta$ (or high temperatures)
$\beta^{-1}=kT>>\omega$ both these integrals tend to
$$4\pi^2{kT\over\hbar}$$

Summing up, we get that the constant $\hbox{Re}\, (g|g)^{\pm}_{\omega}$
for the case of high temperature  will be equal to
${kT\over\hbar}$
up to multiplication by a dimensionless constant depending on the model.

This means that every degree of freeedom that energetically admissible
($kT>>\omega$) and not forbidden by the model
($|\langle\nu|D|\mu\rangle|^2\ne 0$) gives the term of order
${kT\over\hbar}$ in the exponent for dumping of off--diagonal matrix
elements.

The off--diagonal matrix element will dump as $\exp(-tN{kT\over\hbar})$,
where $N$ is the number of degrees of freedom (that for a generic
system can be taken proportional to the number of particles).
Off--diagonal matrix elements describe the quantum interference.
Our result for the dumping of off--diagonal matrix elements (\ref{collapse})
gives us a possibility to distinguish between microscopic systems
(where quantum effects such as quantum interference are important)
and macroscopic system which can be described by classical mechanics.
The macroscopic system is a system satisfying
\begin{equation}\label{muchmore}
N{kT\over\hbar} >>1
\end{equation}
Actually the value $N{kT\over\hbar}$ is of dimension of $t^{-1}$,
and (\ref{muchmore}) means that $\left(N{kT\over\hbar}\right)^{-1}$
is much less than the time of observation.

In the last section of the present paper we will illustrate the collapse
phenomenon (\ref{collapse}) using the quantum extension of the Glauber
dynamics for a system of spins.

We see that the stochastic limit predicts the collapse of a quantum state
into a classical state and, moreover, allows us to estimate the velocity
of the collapse (\ref{collapse}). One can consider (\ref{collapse})
as a more detailed formulation of the Fermi golden rule:
the Fermi golden rule predicts exponential decay of quantum states;
formula (\ref{collapse}) also relates the speed of the decay to the
dimensions (number of particles) of the sstem.

Consider now the system density matrix
$\hat\rho_S\in{\cal C}$, where ${\cal C}$ is the algebra generated by
the spectral projections of the system Hamiltonian $H_S$, and consider
the master equation (\ref{generic}) (we consider the generic case).
We will find
that the evolution defined by this master equation will conserve the algebra
${\cal C}$ and therefore will be a classical evolution. We will show that
this classical evolution in fact describes quantum phenomena.

For $\hat\rho_{S,t}\in{\cal C}$
we define the evolved density matrix of the system
$$
\hat\rho_{S,t}=\sum_\sigma\rho(\sigma,t)
|\sigma\rangle\langle\sigma|
$$
For this density matrix the master equation (\ref{mas})
takes the form
\begin{equation}\label{equforrho}
{d\over dt}\rho(\sigma,t)=
\sum_{\sigma'}\biggl(
\rho(\sigma',t)\left(
2\hbox{Re }(g|g)^-_{\sigma'\sigma} |\langle\sigma|D|\sigma'\rangle|^2
+2\hbox{Re }(g|g)^+_{\sigma\sigma'}|\langle\sigma'|D|\sigma\rangle|^2\right)
-
$$
$$
-\rho(\sigma,t)\left(
2\hbox{Re }(g|g)^+_{\sigma'\sigma}|\langle\sigma|D|\sigma'\rangle|^2
+ 2\hbox{Re }(g|g)^-_{\sigma\sigma'}|\langle\sigma'|D|\sigma\rangle|^2
\right)\biggr)
\end{equation}
Let us note that if $\rho(\sigma,t)$ satisfies the detailed balance
condition
\begin{equation}\label{equlibrium}
\rho(\sigma,t)2\hbox{Re }(g|g)^-_{\sigma\sigma'}
=\rho(\sigma',t)2\hbox{Re }(g|g)^+_{\sigma\sigma'}
\end{equation}
then $\rho(\sigma,t)$ is the stationary solution for (\ref{equforrho}).

Let us investigate (\ref{equforrho}), (\ref{equlibrium})
for the equilibrium state of the field. In this case
$$
2\hbox{Re }(g|g)^-_{\sigma\sigma'}=
2\pi\int dk\, |{g}(k)|^2
\delta(\omega(k)+\varepsilon_{\sigma'}-\varepsilon_{\sigma})
{1\over{1-e^{-\beta\omega(k)}}}=
$$
$$
=2\pi\int dk\, |{g}(k)|^2
\delta(\omega(k)+\varepsilon_{\sigma'}-\varepsilon_{\sigma})
{1\over{1-e^{-\beta(\varepsilon_{\sigma}-\varepsilon_{\sigma'})}}}=
{C_{\sigma\sigma'}\over{1-e^{-\beta
(\varepsilon_{\sigma}-\varepsilon_{\sigma'})}}}
$$
$$
2\hbox{Re }(g|g)^+_{\sigma\sigma'}=
{C_{\sigma\sigma'}\over{e^{\beta
(\varepsilon_{\sigma}-\varepsilon_{\sigma'})}-1}}
$$

The equation (\ref{equforrho}) takes the form
\begin{equation}\label{equiequ}
{d\over dt}\rho(\sigma,t)e^{\beta \varepsilon_{\sigma}}=
\sum_{\sigma'}
{C_{\sigma\sigma'}|\langle\sigma'|D|\sigma\rangle|^2-
C_{\sigma'\sigma}|\langle\sigma|D|\sigma'\rangle|^2
\over{1-e^{-\beta(\varepsilon_{\sigma}-\varepsilon_{\sigma'})}}}
\left(
\rho(\sigma',t)e^{\beta \varepsilon_{\sigma'}}-
\rho(\sigma,t)e^{\beta \varepsilon_{\sigma}}\right)
\end{equation}
Let us note that $C_{\sigma\sigma'}$ are non--zero
(and therefore positive) only if denominators in
(\ref{equiequ}) are positive and $C_{\sigma'\sigma}$ are non--zero
only if the corresponding denominators are negative.

If the system possesses decoherence then $C_{\sigma\sigma'}$,
$C_{\sigma'\sigma}$ are non--zero and the solution of equation
(\ref{equiequ}) for $t\to\infty$ tends to the stationary solution
given by the detailed balance condition (\ref{equlibrium})
$$
{\rho(\sigma,t)\over{1-e^{-\beta
(\varepsilon_{\sigma}-\varepsilon_{\sigma'})}}}=
{\rho(\sigma',t)\over{e^{\beta
(\varepsilon_{\sigma}-\varepsilon_{\sigma'})}-1}}
$$
or
$$
\rho(\sigma,t)e^{\beta \varepsilon_{\sigma}}=
\rho(\sigma',t)e^{\beta \varepsilon_{\sigma'}}
$$
This means that the stationary solution (\ref{equlibrium}) of
(\ref{equforrho}) describes the equilibrium state of the system
$$
\rho(\sigma,t)={e^{-\beta \varepsilon_{\sigma}}\over \sum_{\sigma'}
e^{-\beta \varepsilon_{\sigma'}}}
$$
For a system with decoherence the density matrix will tend, as $t\to\infty$,
to the stationary solution (\ref{equlibrium}) of (\ref{equforrho}).
In particular, as $t\to\infty$, the density matrix collapses to
the classical Gibbs distribution.

The phenomenon of a collapse of a quantum state into a classical state
is connected with the quantum measurement procedure. The quantum
uncertainty will be concentrated at the degrees of freedom of the quantum
field and vanishes after the averaging procedure. One can speculate that the
collapse of the wave function is a property of open quantum systems: we can
observe the collapse of the wave function of the system averaging over the
degrees of freedom of the reservoir interacting with the system.  Usually
the
collapse of a wave function is interpreted as a projection onto a
classical state (the von Neumann interpretation).
The picture emerging from our considearations is more general: the
collapse is a result of the unitary quantum evolution and conditional
expectation (averaging over the degrees of freedom of quantum field).
This is a generalization of the projection: it is easy to see
that every projection $P$ generates a (non identity preserving) conditional
expectation $E_P(X)=PXP$, more generally a set of projections $P_i$
generates the conditional expectation
$$
\sum_i \alpha_i E_{P_i},\qquad \alpha_i\ge 0
$$
but not every conditional expectation could be given in this way.

We have found the effect of the collapse of density matrix for
$\rho(t)=\langle U_t\rho U^*_t\rangle$, where
$U_t=\lim_{\lambda\to 0}e^{it H_0}e^{-it H}$ is the stochastic limit of
interacting evolution. The same effect of collapse will be valid
for the limit of the full evolution $e^{-it H}$, because the full evolution
is the composition of interacting and free evolution. The free evolution
leaves invariant the elements of diagonal subalgebra and multiplies the
considered above nondiagonal element $|\sigma'\rangle\langle\sigma|$ by
the oscillating factor
$e^{it(\varepsilon_{\sigma'}-\varepsilon_{\sigma})}$. Therefore
for the full evolution we get the additional oscillating factor,
and the collapse phenomenon will survive.

\section{Control of coherence}

In this section we generalize the approach of \cite{[4]}
and investigate different regimes of qualitative behavior
for the considered model.

The master equation (\ref{equforrho}) at first sight looks completely
classical. In the present paper we derived this equation using
quantum arguments. Now we will show that (\ref{equforrho}) in fact
describes a quantum behavior.
To show this we consider the following example.

Let us rewrite (\ref{equforrho})
using the particular form (\ref{13}) of $(g|g)^{\pm}$.
Using (\ref{12}), (\ref{13}) we get
\begin{equation}\label{23}
{d\over dt}\rho(\sigma,t)=
\sum_{\sigma'} 2\pi \int dk\, |g(k)|^2
\biggl((N(k)+1)
$$
$$
\left(
\rho(\sigma',t)
\delta(\omega(k)+\varepsilon_{\sigma}-\varepsilon_{\sigma'})
|\langle\sigma|D|\sigma'\rangle|^2-
\rho(\sigma,t)
\delta(\omega(k)+\varepsilon_{\sigma'}-\varepsilon_{\sigma})
|\langle\sigma'|D|\sigma\rangle|^2
\right)+
$$
$$
+N(k)\left(
\rho(\sigma',t)
\delta(\omega(k)+\varepsilon_{\sigma'}-\varepsilon_{\sigma})
|\langle\sigma'|D|\sigma\rangle|^2
-\rho(\sigma,t)
\delta(\omega(k)+\varepsilon_{\sigma}-\varepsilon_{\sigma'})
|\langle\sigma|D|\sigma'\rangle|^2
\right)
\biggr)
\end{equation}
The first term (integrated with $N(k)+1$) on the RHS of this equation
describes the emission of bosons and the second term (integrated with
$N(k)$) describes the absorption of bosons.
For the emission term the part with $N(k)$ describes the induced emission
and
the part with $1$ the spontaneous emission of bosons.

Let us note that the Einstein relation for probabilities
of emission and absorption of bosons with quantum number $k$
$$
{\hbox{probability of emission}\over \hbox{probability of absorption}}=
{N(k)+1\over N(k)}
$$
is satisfied in the stochastic approximation.

The formula (\ref{23}) describes a macroscopic quantum effect.
To show this let us take the spectrum of a system Hamiltonian
(the set of system states $\Sigma=\{\sigma\}$) as follows: let
$\Sigma$ contain two groups $\Sigma_1$ and $\Sigma_2$
of states with the energy gap between
these groups (or, for simplicity, two states $\sigma_1$ and $\sigma_2$
with $\varepsilon_{\sigma_2}>\varepsilon_{\sigma_1}$).
This type of Hamiltonian was considered in different models
of quantum optics, see for review \cite{WaMi}
(for the case of two states we get the spin--boson Hamiltonian
investigated  in \cite{[4]} using the stochastic limit).
Let the state $\langle\cdot\rangle$ of the bosonic field be taken in such
a way that the density $N(k)$, of quanta of the bosonic field, has support
in a set of momentum variables $k$ such that
\begin{equation}\label{nobosons}
0<\omega(k)< \omega_0< |\varepsilon_{\sigma_1}-\varepsilon_{\sigma_2}|,
\qquad k\in \hbox{supp}\, N(k)
\end{equation}
This means that high--energetic bosons are absent.
It is natural to consider the state $\langle\cdot\rangle$
as a sum of equilibrium state at temperature $\beta^{-1}$
and non--equilibrium part. Therefore the density $N(k)$ will
be non--zero for small $k$ because the equilibrium state satisfies
this property.

Under the considered assumption (\ref{nobosons})
the integral of $\delta$--function
$\delta(\omega(k)+\varepsilon_{\sigma_1}-\varepsilon_{\sigma_2})$ with
$N(k)$
in (\ref{23}) equals identically to zero.
Therefore the RHS of (\ref{23}) will be equal to
$$
\sum_{\sigma'}
2\pi\int dk\, |g(k)|^2
\biggl(
\rho(\sigma',t)
\delta(\omega(k)+\varepsilon_{\sigma}-\varepsilon_{\sigma'})
|\langle\sigma|D|\sigma'\rangle|^2-
$$
$$
-\rho(\sigma,t)
\delta(\omega(k)+\varepsilon_{\sigma'}-\varepsilon_{\sigma})
|\langle\sigma'|D|\sigma\rangle|^2
\biggr)
$$
It is natural to consider this value
(corresponding to the spontaneous emission of bosons by the system)
as small with respect to
the induced emission (for $N(k)>>1$).
In this case the density matrix $\rho(\sigma,t)$ will be
almost constant in time.
This is an effect of conservation of quantum coherence:
in the absence of bosons with the energy $\omega(k)$ equal to
$\varepsilon_{\sigma_1}-\varepsilon_{\sigma_2}$
the system cannot jump between
the states $\sigma_1$ and $\sigma_2$ (or, at least,
this transition is very slow), because in the stochastic limit
such jump corresponds to quantum white noise that must be on a mass shell.

At the same time, the transitions between states inside the groups
$\Sigma_1$ and $\Sigma_2$ are not forbidden by (\ref{nobosons}),
because these transitions are connected with the soft bosons
(with small $k$) that are present in the equilibrium part of
$\langle\cdot\rangle$.
In the above assumptions equation (\ref{23})
describes the transition of the system to intermediate equilibrium,
where the transitions between groups of states $\Sigma_1$ and $\Sigma_2$
are forbidden.

If the state $\langle\cdot\rangle$ does not satisfy the property
(\ref{nobosons}), then the system undergoes fast transitions between
states $\sigma_1$ and $\sigma_2$.
We can switch on such a transition by switching on the bosons
with the frequency
$\omega(k)=\varepsilon_{\sigma_2}-\varepsilon_{\sigma_1}$.

In conclusion: equation (\ref{23}) describes a macroscopic quantum effect
controlled by the distribution
of bosons $N(k)$ which can be physically controlled for example by
filtering.

\section{The Glauber dynamics}

In the present section we apply the master equation (\ref{equforrho})
to the derivation of the quantum extension of the classical Glauber
dynamics.
The Glauber
dynamics is a dynamics for a spin lattice with nearest neighbor
interaction, see \cite{Glauber}, \cite{Kawasaki}.
We will prove that the Glauber dynamics can be
considered as a dynamics generated by the master equation of the type
(\ref{equforrho}) derived from a stochastic limit for a quantum spin system
interacting with a bosonic quantum field.

We take the bosonic reservoir space ${\cal F}$ corresponding to
the bosonic equilibrium state at temperature $\beta^{-1}$.
Thus the reservoir state is
Gaussian with mean zero and correlations given by
$$
\langle a^{*}(k)a(k')\rangle=
{1\over{e^{\beta\omega(k)}-1}}\delta(k-k')
$$

For simplicity we only consider the case of a one dimensional spin lattice,
but our considerations extend without any change to multi--dimensional
spin lattices.

The spin variables are labeled by integer numbers $Z$, and,
for each finite subset $\Lambda\subseteq Z$
with cardinality $|\Lambda|$,
the system Hilbert space is
$$
{\cal H}_S={\cal H}_{\Lambda}=\otimes_{r\in\Lambda} C^2
$$
and the system Hamiltonian has the form
$$
H_S=H_{\Lambda}=-{1\over2}\,\sum_{r,s\in\Lambda}J_{rs}\sigma^z_r\sigma^z_s
$$
where $\sigma^x_r$, $\sigma^y_r$, $\sigma^z_r$ are Pauli matrices
$(r\in\Lambda)$ at the $r$-th site in the tensor product
$$
\sigma^i_r=1\otimes\cdots\otimes 1\otimes
\sigma^i\otimes 1\otimes\cdots\otimes 1
$$
For any $r$, $s\in\Lambda$
$$
J_{rs}=J_{sr}\in R, \qquad J_{rr}=0
$$
We consider for simplicity the system Hamiltonian that describes the
interaction of spin with the nearest neighbors (Ising model):
$$
J_{rs}=J_{r,r+1}
$$

The interaction  Hamiltonian $H_I$ (acting in ${\cal H}_S\otimes {\cal F}$)
has the form
$$
H_I=\sum_{r\in
\Lambda}\sigma^x_r\otimes\psi(g_r), \quad \psi(g)=A(g)+A^{*}(g),\quad
A(g)=\int dk \, \overline{g}(k)a(k),
$$
where $\psi$ is a field operator,
$A(g)$ is a smeared quantum field with cutoff function (form factor) $g(k)$.

The eigenvectors $|\sigma\rangle$ of the system Hamiltonian $H_S$ can be
labeled by spin configurations $\sigma$ (sequences of $\pm 1$),
which label the natural basis in ${\cal H}_S$ consisting of
tensor products of eigenvectors of $\sigma^z_r$
(spin up and spin down vectors $|\varepsilon_{r}\rangle$,
corresponding to eigenvalues $\varepsilon_{r}=\pm 1$)
$$
|\sigma\rangle=\otimes_{r\in\Lambda}|\varepsilon_{r}\rangle
$$
In the present section we denote $\varepsilon_{r}$ the energy of the spin
at site $r$, and denote as $E(\sigma)$ the energy of the spin configuration
$\sigma$
$$
E(\sigma)=-{1\over2}\,\sum_{r,s\in\Lambda}J_{rs}
\varepsilon_r\varepsilon_s
$$

The action of the operator $\sigma^x_r$ on the spin configuration $\sigma$
is defined using the action of $\sigma^x_r$ on the corresponding eigenvector
$|\sigma\rangle$: so the operator $\sigma^x_r$ flips the spin at the $r$--th
site in the sequence $\sigma$ (i.e. it maps the vector
$|\varepsilon_{r}\rangle$ in the tensor product into the vector
$|-\varepsilon_{r}\rangle$).
From the form of $H_S$ and $H_I$ it follows that, in (\ref{equforrho}),
the matrix element $\langle\sigma|D|\sigma'\rangle$ of any two eigenvectors,
corresponding to the spin configurations $\sigma$, $\sigma'$,
will be non--zero only if the configurations
$\sigma$, $\sigma'$ differ exactly at one site.
If the configurations $\sigma$, $\sigma'$ differ exactly at one site
then $\langle\sigma|D|\sigma'\rangle=1$.

The (classical) Glauber dynamics will be given by the master
equation for the density matrix laying in the algebra of spectral
projections of the system Hamiltonian (\ref{equforrho})
\begin{equation}\label{glauber}
{d\over dt}\rho(\sigma,t)=
\sum_{r\in\Lambda}\biggl(
\rho(\sigma^x_r\sigma,t)\left(
2\hbox{Re }(g|g)^-_{\sigma^x_r\sigma,\sigma}
+2\hbox{Re }(g|g)^+_{\sigma,\sigma^x_r\sigma}\right)-
$$
$$
-\rho(\sigma,t)\left(
2\hbox{Re }(g|g)^+_{\sigma^x_r\sigma,\sigma}
+ 2\hbox{Re }(g|g)^-_{\sigma,\sigma^x_r\sigma}
\right)\biggr)
\end{equation}
that gives the Glauber dynamics of a system of spins,
see \cite{Glauber}, \cite{Kawasaki}.
Here
\begin{equation}\label{ggminus}
2\hbox{Re }(g|g)^-_{\sigma,\sigma^x_r\sigma}=
2\pi\int dk\, |{g}(k)|^2
\delta(\omega(k)-J_{r-1,r}\varepsilon_{r-1}-J_{r,r+1}\varepsilon_{r+1})
{1\over{1-e^{-\beta\omega(k)}}}
\end{equation}
and analogously all the other $(g|g)^\pm$.

Up to now we have investigated the dynamics for the diagonal part of
the density matrix.
The master equation for the off--diagonal part of the density matrix
(\ref{mas}) will give the quantum extension of the Glauber dynamics.
We consider now this off--diagonal part:
$$
\sum_{\mu\ne\nu} \rho(\mu,\nu,t)|\mu\rangle\langle\nu|
$$
From (\ref{mas}), (\ref{nondiagonal}) we obtain the equation
for the off--diagonal elements of the density matrix
\begin{equation}\label{non-diag-glauber}
{d\over dt} \rho(\mu,\nu,t)=A_{\mu\nu}\rho(\mu,\nu,t)
\end{equation}
\begin{equation}\label{Amunu}
A_{\mu\nu}=\sum_{r\in\Lambda}\biggl(
i\hbox{Im}\, (g|g)^-_{\mu,\sigma^x_r\mu}-
i\hbox{Im}\, (g|g)^-_{\nu,\sigma^x_r\nu} -
i\hbox{Im}\, (g|g)^+_{\sigma^x_r\mu,\mu} +
i\hbox{Im}\, (g|g)^+_{\sigma^x_r\nu,\nu}
$$
$$
-\hbox{Re}\, (g|g)^-_{\mu,\sigma^x_r\mu}
-\hbox{Re}\, (g|g)^-_{\nu,\sigma^x_r\nu}
-\hbox{Re}\, (g|g)^+_{\sigma^x_r\mu,\mu}
-\hbox{Re}\, (g|g)^+_{\sigma^x_r\nu,\nu}
\biggr)
\end{equation}
Equations (\ref{glauber}), (\ref{non-diag-glauber}), (\ref{Amunu})
describe the quantum extension of the classical
Glauber dynamics (\ref{glauber}).
As it was already noted in section 4,
the coefficient $A_{\mu\nu}$ in (\ref{Amunu}) is proportional to
$|\Lambda|$ (the number of particles in the system).
Due to the summation on $r\in\Lambda$
the coefficient $A_{\mu\nu}$ will diverge for large $|\Lambda|$
(the real part of $A_{\mu\nu}$
will tend to $-\infty$). Therefore the density matrix will collapse
to the diagonal subalgebra (the classical distribution function)
very quickly.

Let us consider now the particular case of one dimensional system
with translationally invariant Hamiltonian:
$$
J_{rs}=J_{r,r+1}=J>0
$$
The translationally invariant Hamiltonian does not satisfy the generic non
degeneracy conditions on the system spectrum that we have used in the
derivation of equations (\ref{glauber}), (\ref{non-diag-glauber})
and therefore we cannot apply these equations to describe the dynamics
for this Hamiltonian.

However in the translation invariant one--dimensional case we can
investigate
these equations by direct methods.

In this case the $(g|g)^\pm$, given by (\ref{ggminus}), are non--zero only
if
$\varepsilon_{r-1}=\varepsilon_{r+1}=1$, and we get for (\ref{ggminus})
\begin{equation}\label{C}
2\hbox{Re }(g|g)^-_{\sigma,\sigma^x_r\sigma}=
2\pi\int dk\, |{g}(k)|^2
\delta(\omega(k)-2J)
{1\over{1-e^{-2\beta J}}}= {C\over{1-e^{-2\beta J}}}
\end{equation}

Therefore for one--dimensional translation invariant Hamiltonians
we get for (\ref{glauber}), compare with \cite{Glauber}, \cite{Kawasaki}
\begin{equation}\label{glauber_ti}
{d\over dt}\rho(\sigma,t)=
{C\over{1-e^{-2\beta J}}}\biggl(
\sum_{r\in\Lambda;E(\sigma)>E(\sigma^x_r\sigma)}\left(
e^{-2\beta J}\rho(\sigma^x_r\sigma,t)-\rho(\sigma,t)\right)+
$$
$$
+\sum_{r\in\Lambda;E(\sigma)<E(\sigma^x_r\sigma)}\left(
\rho(\sigma^x_r\sigma,t)-
e^{-2\beta J}\rho(\sigma,t)\right)\biggr)
\end{equation}
The detailed balance stationary solution of (\ref{glauber_ti})
satisfy the following: for two spin configurations $\sigma$,
$\sigma^x_r\sigma$ that differ by the flip of spin at site $r$
the energy of corresponding configurations differ by $2J$.
The expectation $\rho(\mu)$, $\mu=\sigma,\sigma^x_r\sigma$
of configuration with the higher energy will be $e^{-2\beta J}$
times less.

For the off--diagonal part of the density matrix
for the case of one--dimensional translation
invariant Hamiltonian the terms in the imaginary part of (\ref{Amunu})
cancel and using (\ref{C}) we get for (\ref{Amunu})
$$
A_{\mu\nu}=
-\sum_{r\in\Lambda}\left(
{2C\over{1-e^{-2\beta J}}}
+{2C\over{e^{2\beta J}-1}}
\right)=
-2C\sum_{r\in\Lambda}
{{1+e^{-2\beta J}}\over{1-e^{-2\beta J}}}
$$
This sum, over $r$, of equal terms diverges with $|\Lambda|\to\infty$.
Therefore the off--diagonal elements of the density matrix
that satisfy (\ref{non-diag-glauber}) will decay very quickly
and for sufficiently large $t$, $|\Lambda|$ the dynamics of the system
will be given by the classical Glauber dynamics.

For the master equation considered above we used the master equation for
generic (non--degenerate) Hamiltonian. This gives us the Glauber
dynamics. But the translation invariant Hamiltonian is degenerate.
Therefore in the translation invariant case
we will get some generalization of the Glauber dynamics.
To derive this generalization let us consider the general form
(\ref{masterequ}) of the master equation. For the considered spin
system this gives
\begin{equation}\label{newglauber}
\sum_{\mu,\nu} {d\over dt}\rho(\mu,\nu,t)|\mu\rangle\langle\nu|=
\sum_{\mu,\nu}\rho(\mu,\nu,t)\sum_{a,b\in\Lambda}
\biggl(
i\left(
|\mu\rangle\langle P_{E(\nu)}\sigma^x_a P_{\nu-}\sigma^x_b \nu|-
|P_{E(\mu)}\sigma^x_a P_{\mu-}\sigma^x_b\mu\rangle\langle\nu|
\right)+
$$
$$
+{C\over{1-e^{-2\beta J}}}
\biggl(
|P_{E(\mu)-J}\sigma^x_a\mu\rangle\langle P_{E(\nu)-J}\sigma^x_b\nu|-
{1\over 2}\left(
|\mu\rangle\langle P_{E(\nu)}\sigma^x_a P_{E(\nu)-J} \sigma^x_b \nu|
+|P_{E(\mu)}\sigma^x_a P_{E(\mu)-J} \sigma^x_b\mu\rangle\langle\nu|
\right)\biggr)-
$$
$$
-i\left(
|\mu\rangle\langle P_{E(\nu)}\sigma^x_a P_{\nu+}\sigma^x_b\nu|-
|P_{E(\mu)}\sigma^x_a P_{\mu+}\sigma^x_b\mu\rangle\langle\nu|
\right)+
$$
$$
+{C\over{e^{2\beta J}-1}}
\biggl(
|P_{E(\mu)+J}\sigma^x_a\mu\rangle\langle P_{E(\nu)+J}\sigma^x_b\nu|-
{1\over 2}\left(
|\mu\rangle\langle P_{E(\nu)}\sigma^x_a P_{E(\nu)+J} \sigma^x_b \nu|
+|P_{E(\mu)}\sigma^x_a P_{E(\mu)+J} \sigma^x_b\mu\rangle\langle\nu|
\right)\biggr)\biggr)
\end{equation}
Here $C$ is given by (\ref{C}), operator
$P_{E(\mu)}$ is a projector onto the states with the energy $E(\mu)$,
operator $P_{\nu-}$ is given by
$$
P_{\nu-}=
\hbox{ Im }(g|g)^-_{-1}P_{E(\nu)-J}+
\hbox{ Im }(g|g)^-_{0}P_{E(\nu)}+\hbox{ Im }(g|g)^-_{1}P_{E(\nu)+J}
$$
$$
\hbox{ Im }(g|g)^-_{a}=\hbox{P.P.}\, \int dk\, |g(k)|^2 {1\over \omega(k)+aJ}
{1\over 1-e^{-\beta\omega(k)}},
\qquad  a=-1,0,1
$$
For the operator $P_{\nu+}$ we get the analogous expression
$$
P_{\nu+}=
\hbox{ Im }(g|g)^+_{-1}P_{E(\nu)-J}+
\hbox{ Im }(g|g)^+_{0}P_{E(\nu)}+\hbox{ Im }(g|g)^+_{1}P_{E(\nu)+J}
$$
with the coefficients $(g|g)^+_{a}$:
$$
(g|g)^+_{a}=\hbox{P.P.}\, \int dk\, |g(k)|^2 {1\over \omega(k)-aJ}
{1\over e^{\beta\omega(k)}-1},
\qquad  a=-1,0,1
$$
The equation (\ref{newglauber}) gives the quantum generalization
of the Glauber dynamics.
The matrix elements $\rho(\mu,\nu,t)$ of the density matrix corresponding to
the states $\mu$, $\nu$ with different energies will decay quickly.
But for the translation invariant Hamiltonian there exist different
$\mu$, $\nu$ with equal energies. Corresponding matrix element will
decay with the same speed as the diagonal elements of the density matrix.
Moreover one can expect non--ergodic behavior for this model.
Therefore the generalization (\ref{newglauber}) of the Glauber dynamics
is non--trivial.

\subsection{Evolution for subalgebra of local operators}

In this section to compare with the results of \cite{[5]}
we consider the dynamics of spin systems, described in the previous
section, for Hamiltonian with non necessarily finite
set of spins $\Lambda$ but for local observable $X$.

The observable $X$ is local if it belongs to the local algebra,
that is UHF--algebra (uniformly hyperfinite algebra)
$$
{\cal A}= \bigcup_{\Lambda {\tiny\hbox{ is finite }}} {\cal A}_{\Lambda}
$$
where ${\cal A}_{\Lambda}$ is the $*$--algebra generated by the elements
$$
\otimes_i X_i, \qquad X_i= 1 \hbox{ for } i\not\in \Lambda
$$

Consider now the action of $\theta_0$ on local $X$
\begin{equation}\label{localtheta}
\theta_0(X)=\sum_{ij} \sum_{\omega\in F}
\biggl(
-i\hbox{ Im }(g_i|g_j)^-_{\omega}
[X,E_{\omega}^*\left(D_i\right)E_{\omega}\left(D_j\right)]+
i\hbox{ Im }{(g_i|g_j)}^+_{\omega}
[X,E_{\omega}\left(D_i\right)E_{\omega}^*\left(D_j\right)]+
$$
$$+
2\hbox{Re}\,(g_i|g_j)^-_{\omega}
\left(
E_{\omega}^*\left(D_i\right) X E_{\omega}\left(D_j\right)
-{1\over 2}
\{X,E_{\omega}^*\left(D_i\right)E_{\omega}\left(D_j\right)\}\right)+
$$
$$
+2\hbox{Re}\,{(g_i|g_j)}^+_{\omega}
\left(
E_{\omega}\left(D_i\right) X E_{\omega}^*\left(D_j\right)
-{1\over 2}
\{X,E_{\omega}\left(D_i\right)E_{\omega}^*\left(D_j\right)\}\right)\biggr)
\end{equation}

For $E_{\omega}\left(D_i\right)$ we get
\begin{equation}\label{localE}
E_{\omega}\left(D_i\right)=
\sum_{E(r)\in F_{\omega}}
P_{E(r)-\omega} D_i P_{E(r)}=
$$
$$
=1\otimes|\varepsilon_{i-1}\rangle\langle\varepsilon_{i-1}|\otimes
|-\varepsilon_{i}\rangle\langle\varepsilon_{i}|\otimes
|\varepsilon_{i+1}\rangle\langle\varepsilon_{i+1}|\otimes 1+
1\otimes|-\varepsilon_{i-1}\rangle\langle-\varepsilon_{i-1}|\otimes
|\varepsilon_{i}\rangle\langle-\varepsilon_{i}|\otimes
|-\varepsilon_{i+1}\rangle\langle-\varepsilon_{i+1}|\otimes 1;
\end{equation}
with the frequency $\omega$ of the following form
$$
\omega=J_{i-1,i}\varepsilon_{i-1}+J_{i,i+1}\varepsilon_{i+1}
$$

Therefore the operator $E_{\omega}$ given by (\ref{localE})
is local and moreover, corresponding map $\theta_0$ given by
(\ref{localtheta}) maps ${\cal A}$ into itself.

The formula (\ref{localE}) explains the physical meaning of
the operator $E_{\omega}\left(D_i\right)$. For positive $\omega$
this it flips the spin at site $i$
along the direction of the mean field of its neighbors
(for negative $\omega$ it flips the same spin into the opposite direction).

\bigskip

\centerline{\bf Acknowledgments}\smallskip

Sergei Kozyrev is grateful to Luigi Accardi and Centro Vito Volterra where
this work was done for kind hospitality.
The authors are grateful to I.V.Volovich for discussions.
This work was partially supported by INTAS 99--00545 grant.
Sergei Kozyrev was partially supported by RFFI 990100866 grant.

\end{document}